\documentclass[twocolumn,floats,floatfix,aps,pra]{revtex4-2}
\usepackage{amsfonts,amssymb,amsmath}
\usepackage{physics}
\usepackage{color,calc}
\usepackage{bm}
\usepackage{amsbsy,amsmath}
\usepackage{array}
\usepackage{booktabs}
\usepackage{graphicx}
\usepackage{graphics}
\usepackage{eepic,epsfig}

\usepackage[breaklinks,hidelinks,colorlinks=true,linkcolor=blue,citecolor=blue,filecolor=blue,urlcolor=blue]{hyperref}

\usepackage[dvipsnames]{xcolor}
\definecolor{myblue}{RGB}{0, 0, 255}
\usepackage[toc,page]{appendix}

\def\be{ \begin{equation} }
\def\ee{ \end{equation} }
\def\bea{ \begin{eqnarray} }
\def\eea{ \end{eqnarray} }
\def\bse{ \begin{subequations} }
\def\ese{ \end{subequations} }
\def\ba{ \begin{array} }
\def\ea{ \end{array} }
\def\bt{ \begin{tabular} }
\def\et{ \end{tabular} }

\def\i{\,\text{i}}

\def\i{i}

\def\d{\text{d}}

\def\U{\mathbf{U}}
\def\F{\mathbf{F}}

\def\A{\mathcal{A}}

\def\i{{\rm{i}}}
\def\f{{\rm{f}}}
\def\phase{\phi}

\long\def\/*#1*/{}

\begin{document}

\title{Ultrahigh-fidelity composite quantum phase gates}

\author{Hayk L. Gevorgyan\textsuperscript{\hyperref[dag]{\dag},\hyperref[ddag]{\ddag},\hyperref[S]{\S}}}
\author{Nikolay V. Vitanov\textsuperscript{\hyperref[dag]{\dag}}}
\affiliation{\phantomsection\label{dag}{\textsuperscript{\dag}Faculty of Physics, St Kliment Ohridski University of Sofia, 5 James Bourchier blvd, 1164 Sofia, Bulgaria}\\
\phantomsection\label{ddag}{\textsuperscript{\ddag}A. Alikhanyan National Science Laboratory (Yerevan Physics Institute), 2 Alikhanian Brothers st, 0036 Yerevan, Armenia}\\
\phantomsection\label{S}{\textsuperscript{\S}Institute for Physical Research, Armenian National Academy of Sciences, Ashtarak-2, 0203, Armenia}}

\date{\today }

\begin{abstract}
A number of composite pulse (CP) sequences for four basic quantum phase gates --- the Z, S, T and general phase gates --- are presented.
The CP sequences contain up to 18 pulses and can compensate up to eight orders of experimental errors in the pulse amplitude and duration.
The short CP sequences (up to 8 pulses) are calculated analytically and the longer ones numerically.
The results demonstrate the remarkable flexibility of CPs accompanied by extreme accuracy and robustness to errors --- three features that cannot be simultaneously achieved by any other coherent control technique.
These CP sequences, in particular the Z, S and T gates, can be very useful quantum control tools in quantum information applications, because they provide a variety of options to find the optimal balance between ultrahigh fidelity, error range and speed, which may be different in different physical applications.
\end{abstract}

\maketitle



\section{Introduction\label{Sec:intro}}


Phase coherence is of paramount importance in modern quantum information technologies and it is one of the most significant differences between classical and quantum computing \cite{nielsen2000, vandersypen2005, jones2011, chen2006}. 
Phase coherence is controlled by quantum phase gates, such as the Z, S and T gates, which are key elements in any quantum circuit.
Because of the vast number of such gates involved even in moderate quantum circuits their fidelity is of crucial significance for the success of any quantum algorithm.

Among the existing quantum control techniques capable of efficient manipulation of quantum systems, composite pulse (CP) sequences \cite{Levitt1986, Levitt2007} stand out as a very powerful tool which offers a unique combination of accuracy of operations, robustness to experimental errors, flexibility and versatility as it can be adopted and applied to essentially any quantum control task --- a set of features that can only be found in composite pulses.
A composite pulse is actually a sequence of pulses with well defined relative phases which are used as control parameters in order to shape the excitation profile, and generally, the propagator, in essentially any desired manner.

The vast majority of composite pulses are designed to produce complete and partial rotations on the Bloch sphere \cite{Levitt1986, Wimperis1990, Wimperis1991, wimperis, Levitt2007, Torosov2019variable, Torosov2020, gevorgyan2021}.
Among these, a clear distinction exists between the so-called variable and constant rotations. 
Variable rotations start on one of the poles of the Bloch sphere and move the Bloch vector at a particular latitude, i.e. on a particular parallel, without controlling the longitude. Constant rotations do not require a specific initial condition and produce the desired rotation starting at any point on the Bloch sphere. 
In quantum control language, the variable rotations are characterized by well-defined absolute values (i.e. populations) of the propagator elements but not well-defined phases. 
Constant rotations (or phase-distortionless rotations) are characterized by both well-defined populations and phases of the propagator, i.e. the quantum gate.
Obviously, constant rotations are much more demanding to generate, but they are exactly what is required for reliable and scalable quantum computing circuits.
Over the years, variable and constant composite rotations have been demonstrated on multiple occasions in NMR \cite{Levitt1979, Freeman1980, Levitt1982, Levitt1983, Levitt1986, Levitt2007, Wimperis1990, Wimperis1991, wimperis}, trapped ions \cite{Gulde2003, Schmidt-Kaler2003, Timoney2008, Monz2009, Shappert2013, Mount2015, Randall2018, Zarantonello2019}, neutral atoms \cite{Rakreungdet2009, Butts2013, Dunning2014, Berg2015, Zanon-Willette2018}, quantum dots \cite{Wang2012,Kestner2013,Wang2014,Zhang2017,Hickman2013,Eng2015}, doped solids \cite{Schraft2013,Genov2017,Bruns2018,genov2014}, superconducting qubits \cite{SteffenMartinisChuang, Torosov2022}, etc., featuring remarkable accuracy and robustness.
A variation of the composite idea, with the detuning rather than the phase of each constituent pulse in the composite sequence used as the control parameter, has also been proposed and experimentally demonstrated \cite{Ivanov2022}.

Compared to rotations, very few proposals exist for composite phase gates \cite{torosov2014}.
An arbitrary phase shift at an angle $\phi$, being rotation around the $z$ axis, can be implemented by two resonant $\pi$ pulses with an appropriate relative phase between them.
However, resonant driving is prone to errors in the experimental parameters, e.g. the pulse amplitude, duration, and detuning.
Here the phase gates are implemented as sequences of $\pi$ pulses with specific phases. 
Application of composite pulses to quantum phase gates has been discussed in Ref.~\cite{torosov2014}, where composite sequences of $2(2n+1)$ ($n=1,2,\ldots$) pulses have been presented, with the 6-pulse and 10-pulse sequences being the shortest ones. 
In the present paper, we make a step toward filling this gap: we supplement the library of composite pulses with composite sequences of $2n$ ($n=2,3,\ldots$) pulses, which produce arbitrary quantum phase gates, with a focus at the most important ones for quantum information processing: the Z, S, and T gates.
We use analytic approaches and brute-force numerics to derive composite sequences, which achieve error compensation of up to 8th order. 

This paper is organized as follows.
In Sec.~\ref{Ch3:Sec:derivation} we explain the derivation method.
The design and performance of the composite phase gates are presented in Sec.~\ref{Sec:phase_gates}.
Finally, Sec.~\ref{Sec:concl-ch3} presents the conclusions.



\section{SU(2) Approach}\label{Ch3:Sec:derivation}


\subsection{Method of derivation}

Our objective in this article is to construct the quantum phase-shift gate $\F(\phi) = e^{-i (\phi/2) \hat\sigma_z}$, or in matrix form,  
\be\label{F}
\mathbf{F}(\phi) = \mathbf{R}_z(\phi) = \left[ \begin{array}{cc}  e^{-i \phi/2} & 0 \\  0 & e^{i \phi/2} \end{array} \right] := \left[ \begin{array}{cc}  1 & 0 \\  0 & e^{i \phi} \end{array} \right] ,
\ee
which is equal to the standard one up to the unimportant global phase factor $e^{-i \phi/2}$.
For the Z gate we have $\phi=\pi$, for the S gate $\phi=\pi/2$, and for the T gate $\phi=\pi/4$.

The derivation of the robust ultrahigh-fidelity quantum phase gates via composite pulses is similar to the rotation gates in our previous work \cite{gevorgyan2021}. 
Basically, starting from the time-dependent Schr\"odinger equation for a two-state system, one can write down the evolution operator for a single-qubit, which is referred to as Rabi rotation gate 
in experimental quantum computing \cite{chen2006, nielsen2000}, or theta pulse in nuclear magnetic resonance \cite{wimperis},
\be\label{U1}
\U_{\phi} (\A) = \left[ \begin{array}{cc} \cos(\A/2) & -i e^{i\phi} \sin(\A/2) \\ -i e^{-i\phi} \sin(\A/2) & \cos(\A/2)  \end{array}  \right].
\ee
Here $\A=\int_{t_\i}^{t_\f}\Omega(t)\d t$ is the temporal pulse area and $\phi$ stands for the phase of the coupling.
A train of $N$ resonant theta pulses, each with a specific pulse area $\A_k$ and a specific phase $\phase_k$ ($k=1,2,\ldots,N$),
\be\label{design}
(\A_1)_{\phi_1} (\A_2)_{\phi_2} (\A_3)_{\phi_3} \cdots (\A_N)_{\phi_{N}},
\ee
produces the propagator 
\be\label{U^N}
\boldsymbol{\mathcal{U}} = \U_{\phase_{N}}(\A_N) \cdots \U_{\phase_{3}}(\A_3) \U_{\phase_{2}}(\A_2) \U_{\phase_{1}}(\A_1).
\ee
In Eq.~\eqref{U^N} the evolution matrices $\U_{\phase_{k}}(\A_k)$ act chronologically, from right to left, while in Eq.~\eqref{design} the pulses $(\A_k)_{\phi_k}$ are applied from left to right. 

Under the assumption of a single systematic pulse area error $\epsilon$, i.e. when each pulse is replaced by the errant one, $\A_k \rightarrow \A_k (1+\epsilon)$, we can expand the errant composite propagator
\be\label{U^N-err}
\boldsymbol{\mathcal{U}} (\epsilon) = \left[ \begin{array}{cc}  \mathcal{U}_{11}(\epsilon) & \mathcal{U}_{12}(\epsilon) \\  -\mathcal{U}_{12}^{\ast}(\epsilon) & \mathcal{U}_{11}^{\ast}(\epsilon) \end{array} \right] .
\ee
in a Taylor series versus $\epsilon$. Because of the SU(2) symmetry of the errant overall propagator, it suffices to expand only two of its elements, say $\mathcal{U}_{11}(\epsilon)$ and $\mathcal{U}_{12}(\epsilon)$. 
We set their zero-error (nominal) values to the target values,
\be\label{eq-0-phase}
\mathcal{U}_{11}(0) = e^{-i \phi/2},\quad \mathcal{U}_{12}(0) = 0,
\ee
and we set to zero as many of their derivatives with respect to $\epsilon$, in the increasing order, as possible,
\be\label{eq-m-phase}
\mathcal{U}^{(m)}_{11}(0) = 0,\quad \mathcal{U}^{(m)}_{12}(0) = 0, \quad (m=1,2,\ldots, n),
\ee
where $ \mathcal{U}^{(m)}_{jl} = \partial_\epsilon^m  \mathcal{U}_{jl}$ denotes the $m$th derivative of $\mathcal{U}_{jl}$ with respect to $\epsilon$.
The largest derivative order $n$ satisfying Eqs.~\eqref{eq-m-phase} gives the order of the error compensation $O(\epsilon^n)$.

Equations \eqref{eq-0-phase} and \eqref{eq-m-phase} generate a system of $2(n+1)$ algebraic equations for the nominal pulse areas $A_k$ and the composite phases $\phi_k$ ($k=1,2,\ldots,N$). 
The equations are complex-valued and generally we have to solve $4(n+1)$ equations with the $2N$ free parameters (nominal pulse areas and phases). 
Equation \eqref{eq-0-phase} alone can be satisfied at least by two $\pi$ pulses, 
\be
 \pi_\nu \pi_{\nu + \pi - \phi/2} , \\
\ee
with the propagator
\be\label{F-two}
\mathbf{F}(\phi) = \U_{\nu + \pi - \phi/2} (\pi) \U_{\nu} (\pi) . 
\ee
For the Z, S and T gates we have the sequences (setting $\nu=0$)
\bse
\begin{align}
{Z2} &= \pi_0 \pi_{\frac12 \pi} , \\
{S2} &= \pi_0 \pi_{\frac34 \pi} , \\
{T2} &= \pi_0 \pi_{\frac78 \pi} , 
\end{align}
\ese
Taking into account this fact, and because of the normalization condition $|\mathcal{U}_{11}|^2 + |\mathcal{U}_{12}|^2 = 1$ ($\mathcal{U}_{11}$ and $\mathcal{U}_{12}$ are the complex-valued Cayley-Klein parameters), an error compensation of order $n$ requires a CP sequence of $N=2(n+1)$ $\pi$ pulses.

As stated above, the derivation of the CP sequences requires the solution of Eqs.~\eqref{eq-0-phase} and \eqref{eq-m-phase}.
For composite sequences of a small number of pulses (up to eight $\pi$ pulses), Eqs.~\eqref{eq-0-phase} and the first, second and third pairs of equations ($n=3$) of Eqs.~\eqref{eq-m-phase} can be solved analytically.
For longer pulse sequences, Eqs.~\eqref{eq-m-phase} are solved numerically.
We do this by using standard routines in \textsc{Mathematica}
$^\copyright$.

\subsection{Quantum gate fidelity}

If Eqs.~\eqref{eq-0-phase} and \eqref{eq-m-phase} are satisfied, then the overall propagator can be written as
\be\label{U-epsilon}
\boldsymbol{\mathcal{U}} (\epsilon) = \mathbf{F}(\phi) + O(\epsilon^{n+1}),
\ee
with $\mathbf{F}(\phi) = \boldsymbol{\mathcal{U}} (0)$.
Then the \emph{Frobenius distance fidelity},
\be\label{Frobenius}
\mathcal{F} = 1 - \| \boldsymbol{\mathcal{U}} (\epsilon) - \mathbf{F}(\phi) \|
 = 1 - \sqrt{ \tfrac14 \sum\nolimits_{j,k=1}^2 \left|\mathcal{U}_{jk} - F_{jk} \right|^2 } ,
\ee
is of the same error order $O(\epsilon^{n})$ as the propagator, $\mathcal{F} = 1 - O(\epsilon^{n+1})$.
Another possibility is to use the \emph{trace distance fidelity},
\be\label{trace fidelity}
\mathcal{F}_{\text{T}} = \tfrac12 \text{Tr}\, [ \boldsymbol{\mathcal{U}} (\epsilon) \mathbf{F}(\phi)^\dagger ] .
\ee
As in our previous work \cite{gevorgyan2021}, we will use the Frobenius distance fidelity, because it explicitly includes information about both major and minor diagonal elements, as required for phase-distortionless phase gates.


\subsection{Structure of composite phase gates\label{Ch3-results}}

Based on numerical evidence, we consider CP sequences, which consist of $2(n+1)$ nominal $\pi$ pulses, with asymmetrically ordered phases,
\bse\label{phase-asymmetric}
\begin{align}
& R_{n+1}(\nu)  R_{n+1}(\nu+\pi-\phi/2) , \\
& R_{n+1}(\nu) = 
\pi_{\nu} \pi_{\nu+\phi_1} \pi_{\nu+\phi_2} \cdots \pi_{\nu+\phi_{n}} , \label{phase-asymmetric-str}
\end{align}
\ese
which is equivalent to 
\be\label{phase-asymmetric-2}
R_{n+1}(\nu + \pi+ \phi/2)  R_{n+1}(\nu) . 
\ee
%
These sequences generalize the initial two-pulse sequence \eqref{F-two} and have similar design. 
Due to this specific structure of composite phases, Eqs.~\eqref{eq-0-phase} are satisfied, all odd-order derivatives $\mathcal{U}^{(2k+1)}_{11}(0)$ of the major-diagonal elements in Eq.~\eqref{eq-m-phase} vanish, and so do all even-order derivatives $\mathcal{U}^{(2k)}_{12}(0)$ of the minor-diagonal elements. 
This facilitates reaching the compensation order $n$, which is the maximum number for which the derivatives of \textit{all} major-diagonal and minor-diagonal elements vanish. 
This can be obtained with a suitable choice of the available composite phases in Eq.~\eqref{phase-asymmetric}.

From the infinite number of solutions, we choose those of the type \eqref{phase-asymmetric} and with a free parameter $\nu = 0$, since only the choice of the relative phases $\phi_1, \phi_2, \ldots, \phi_n$ is of importance. 
Henceforth, we use the design    
\bse\label{phase-asymmetric-0}
\begin{align}
& R_{n+1}  R_{n+1}(\pi-\phi/2), \\
& R_{n+1} = 
\pi_{0} \pi_{\phi_1} \pi_{\phi_2} \cdots \pi_{\phi_{n}}, 
\\
& R_{n+1}(\pi-\phi/2) = 
\pi_{\pi-\frac{1}{2}\phi} \pi_{\phi_1 + \pi-\frac{1}{2}\phi} \cdots \pi_{\phi_{n} + \pi-\frac{1}{2}\phi}.
\end{align}
\ese
Other possible but equivalent solutions can be obtained by choosing an arbitrary parameter $\nu$ in \eqref{phase-asymmetric} or/and by using the type \eqref{phase-asymmetric-2}.

\section{Composite phase gates}\label{Sec:phase_gates}

As it is well known, such a gate can be produced by two resonant $\pi$ pulses, see Eq.~\eqref{F-two} with $\nu=0$.
The propagator elements of the overall propagator read
\bse
\begin{align}\label{U-two-pi}
 \mathcal{U}_{11}(\epsilon) &= e^{-i \phi/2}\cos^2(\pi \epsilon/2) + \sin^2(\pi \epsilon/2), \\
 \mathcal{U}_{12}(\epsilon) &= \tfrac12 i (1 - e^{-i \phi/2}) \sin(\pi \epsilon) ,
 \end{align}
\ese
where $\epsilon$ is the pulse area error.
The Frobenius distance fidelity \eqref{Frobenius} reads 
\be \label{F-2-distance}
\mathcal{F} = 1 - \sqrt{2} \left| \sin \frac{\pi  \epsilon }{2} \right| \left| \sin \frac{\phi}{4} \right|,
\ee
and it has 0th order error compensation, $\mathcal{F} = 1 + O(\epsilon)$.
For comparison, the trace fidelity is
\be \label{F-2-trace}
\mathcal{F}_T = 1 - 2  \sin^2 \frac{\pi  \epsilon }{2} \sin^2 \frac{\phi }{4},
\ee
which has 1st order error compensation, $\mathcal{F} = 1 + O(\epsilon^2)$
Obviously the error of the Frobenius distance fidelity \eqref{F-2-distance} is far greater than the value of the error of the trace fidelity \eqref{F-2-trace}.
%
Below we consider longer sequences, in the increasing order of error compensation.

\begin{figure}[t]
\bt{r}
\includegraphics[width=0.94\columnwidth]{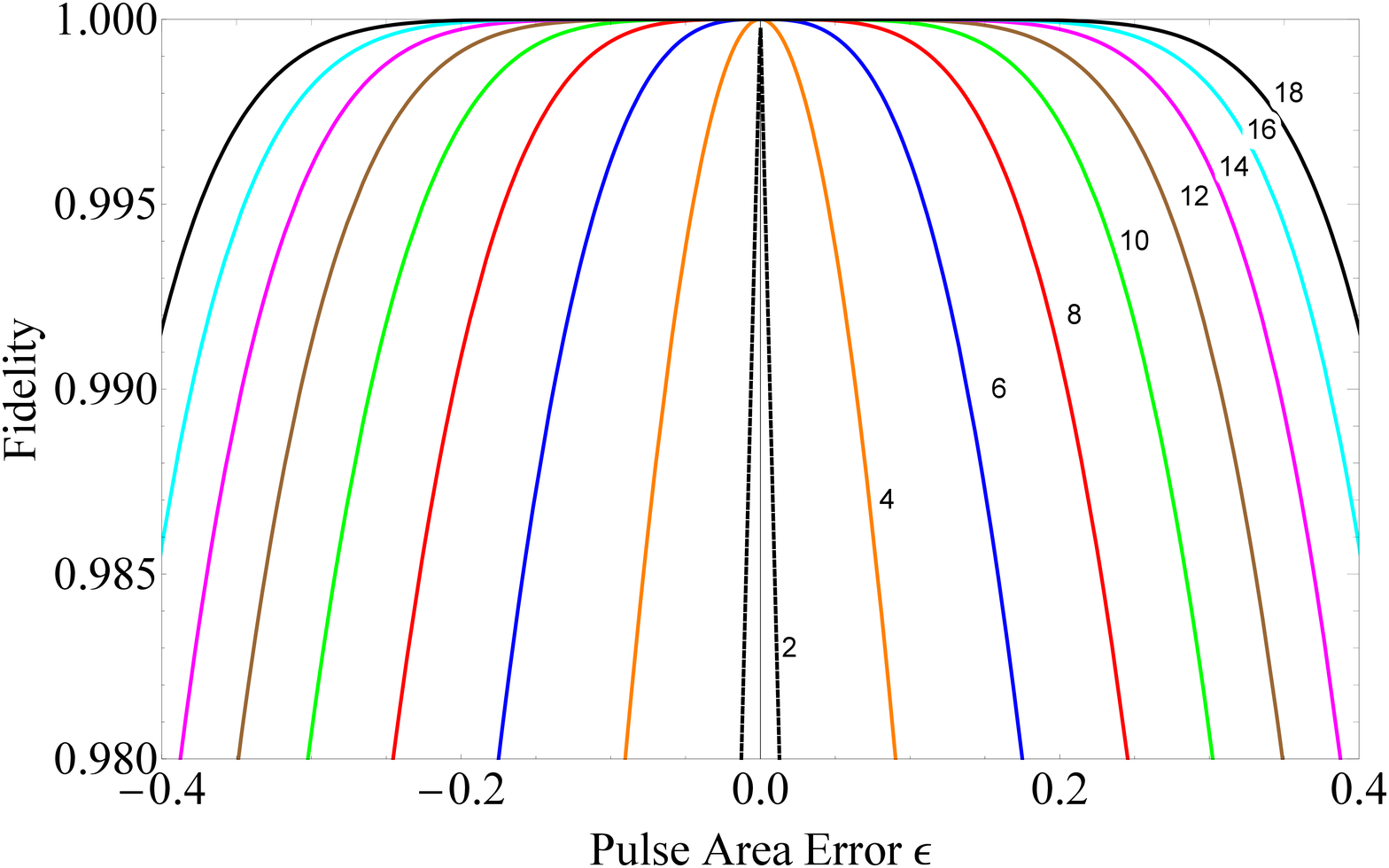} \\ \\
\includegraphics[width=0.94\columnwidth]{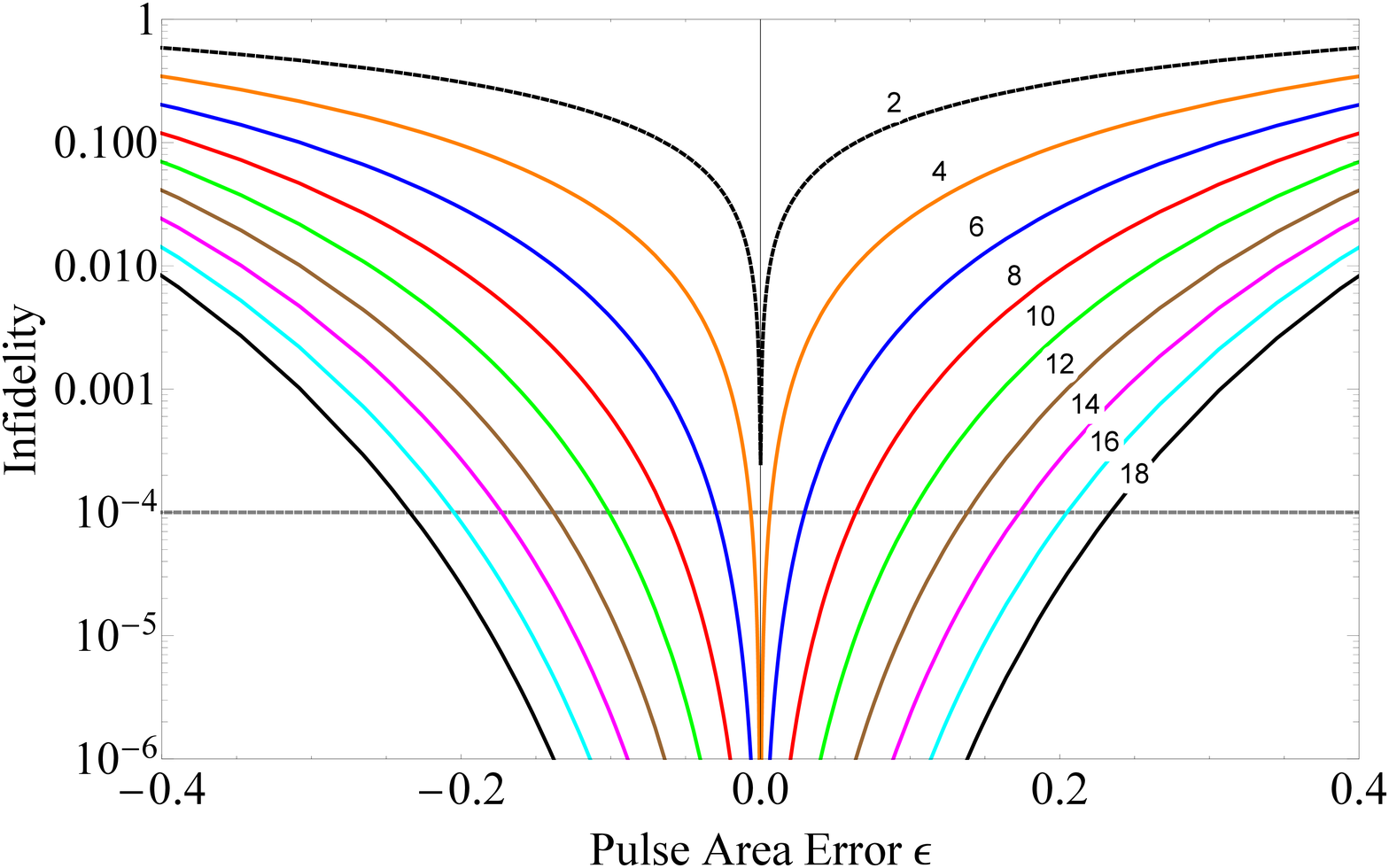}
\et
\caption{
Frobenius distance fidelity $\mathcal{F}$ (top) and infidelity $1-\mathcal{F}$ (bottom) of composite Z gates.
The infidelity is in logarithmic scale in order to better visualize the high-fidelity (low-infidelity) range.
The numbers $N$ on the curves refer to CP sequences Z$N$ listed in the Table~\ref{Table:Z}.
}
\label{fig:Z}
\end{figure}

\begin{figure}[tb]
\bt{r}
\includegraphics[width=0.94\columnwidth]{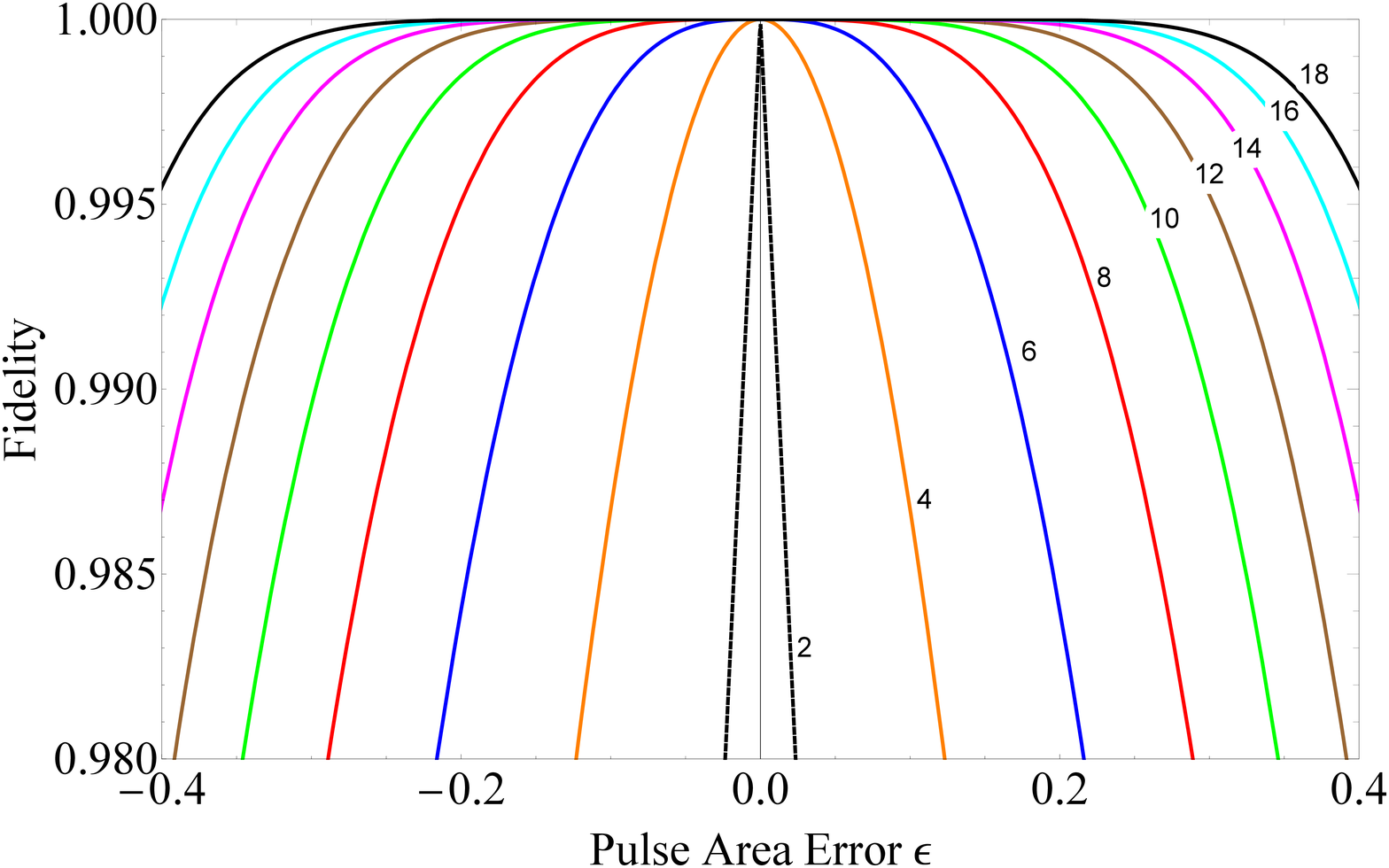} \\ \\
\includegraphics[width=0.94\columnwidth]{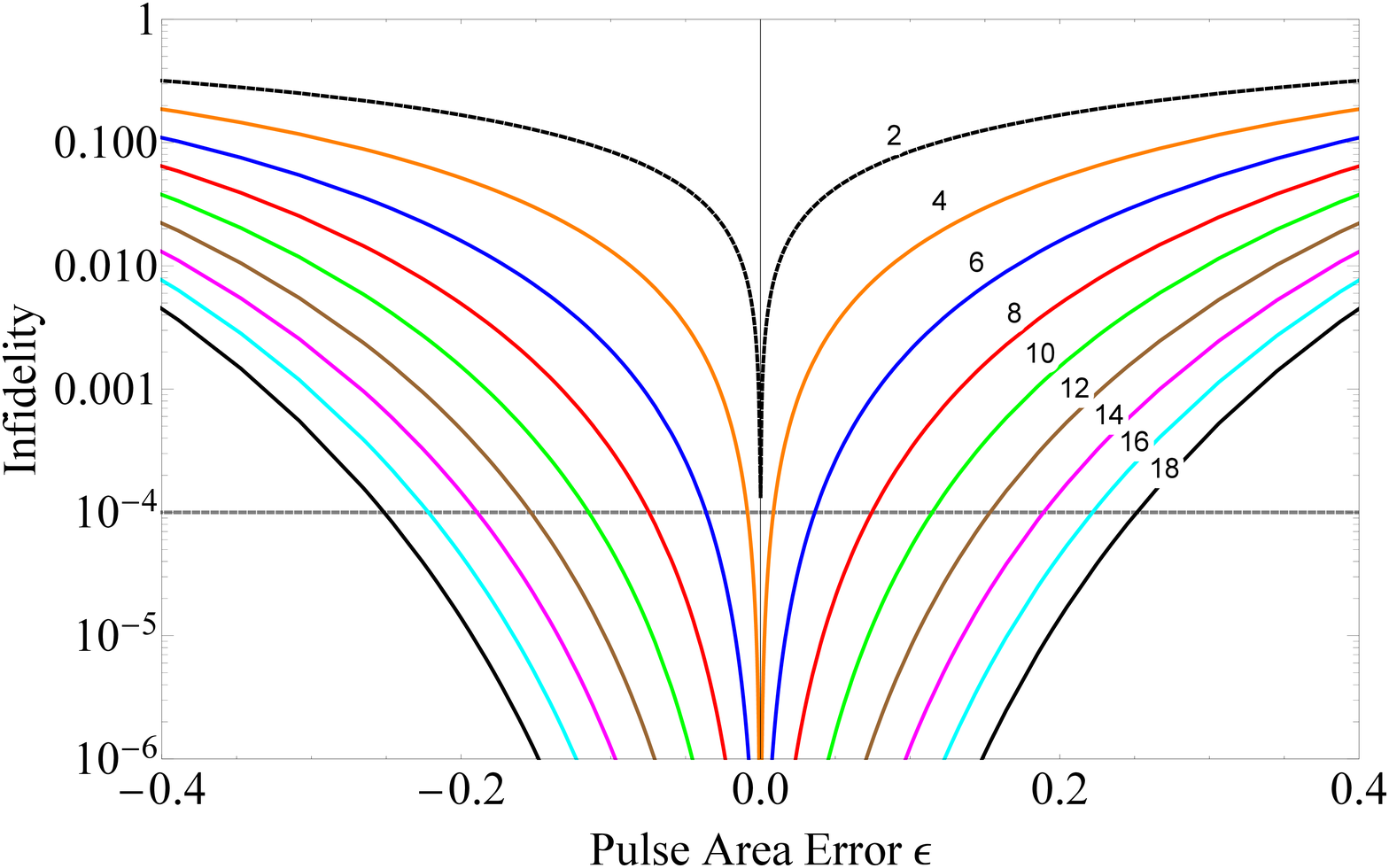}
\et
\caption{
Frobenius distance fidelity $\mathcal{F}$ (top) and infidelity $1-\mathcal{F}$ (bottom) of composite S gates.
The infidelity is in logarithmic scale in order to better visualize the high-fidelity (low-infidelity) range.
The numbers $N$ on the curves refer to CP sequences S$N$ listed in the Table~\ref{Table:S}.
}
\label{fig:S}
\end{figure}

\begin{figure}[t]
\bt{r}
\includegraphics[width=0.94\columnwidth]{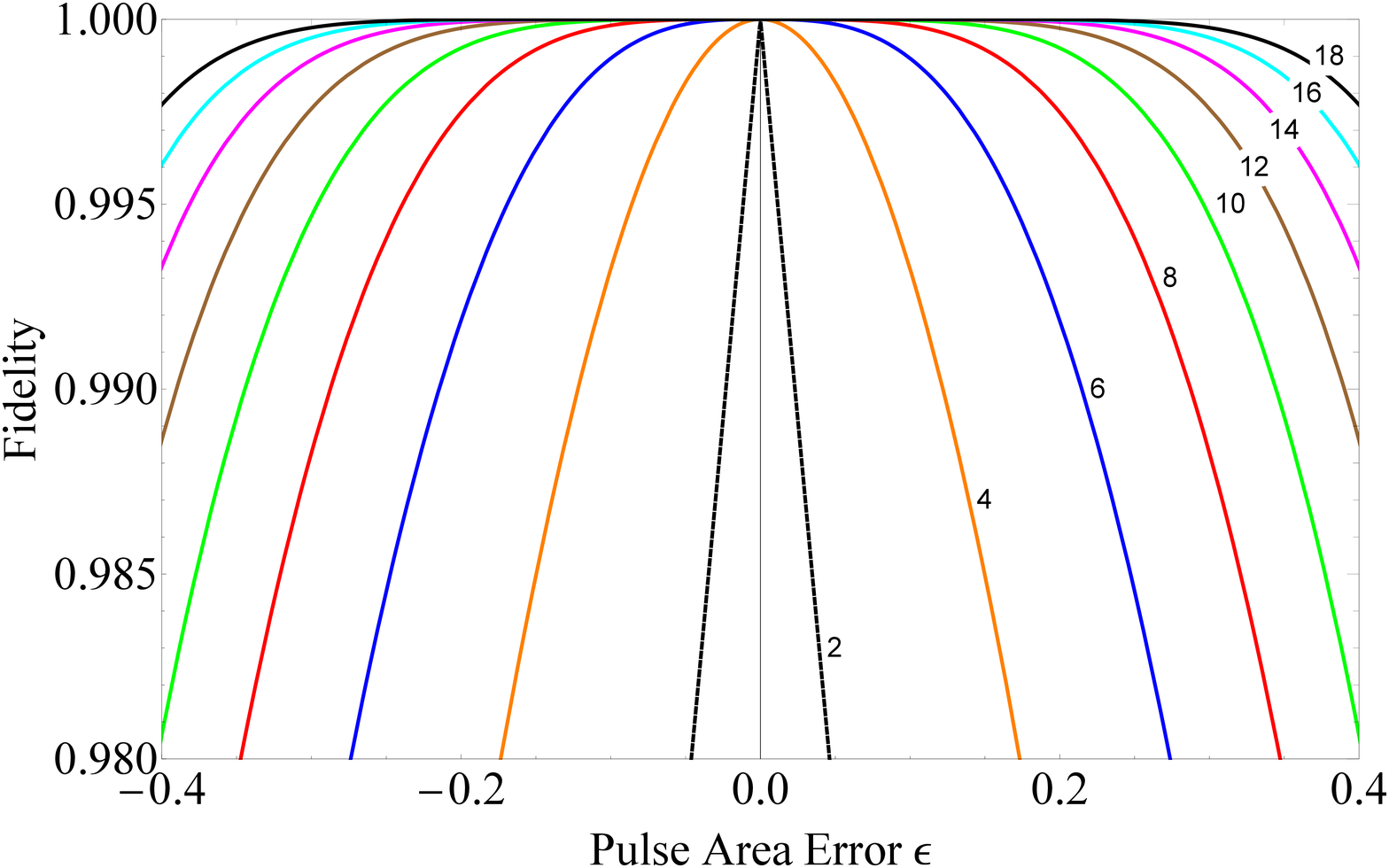} \\ \\
\includegraphics[width=0.94\columnwidth]{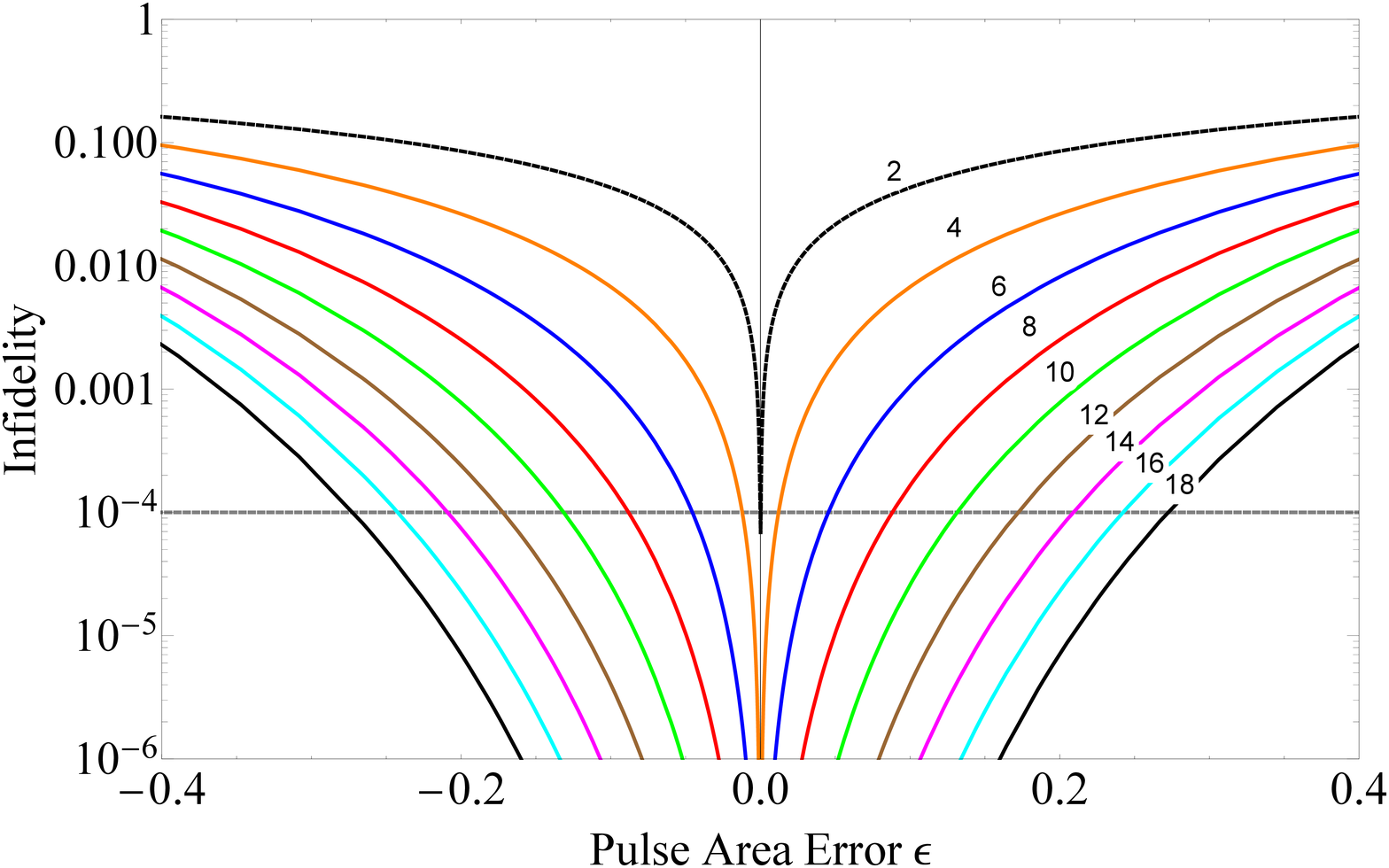}
\et
\caption{
Frobenius distance fidelity $\mathcal{F}$ (top) and infidelity $1-\mathcal{F}$ (bottom) of composite T gates.
The infidelity is in logarithmic scale in order to better visualize the high-fidelity (low-infidelity) range.
The numbers $N$ on the curves refer to CP sequences T$N$ listed in the Table~\ref{Table:T}.
}
\label{fig:T}
\end{figure}

\subsection{First-order error compensation}\label{Sec:phase-4}

The careful analysis of Eqs.~\eqref{eq-0-phase} and \eqref{eq-m-phase} shows that the shortest possible CP which can compensate first-order errors (both in major and minor diagonal elements) consists of four pulses, each with a nominal pulse area of $\pi$, and asymmetric phases, with the structure similar to the two-pulse case, 
\be\label{phase4}
\pi_{0} \pi_{\phi_1} \pi_{\pi - \frac12 \phi} \pi_{\phi_1 + \pi - \frac12 \phi}.
\ee
Solving Eqs.~\eqref{eq-0-phase} along with Eqs.~\eqref{eq-m-phase} for the first derivatives gives four solutions for the phases,
\bse\label{phase4-two}
\begin{align}
& \pi_{0} \pi_{- \frac14 \phi} \pi_{\pi - \frac12 \phi} \pi_{\pi - \frac34 \phi}, \label{phase-four-0}\\
& \pi_{0} \pi_{\pi - \frac14 \phi} \pi_{\pi - \frac12 \phi} \pi_{- \frac34 \phi},\\
& \pi_{\frac14 \phi} \pi_{0} \pi_{\pi - \frac14 \phi} \pi_{\pi - \frac12 \phi},\\
& \pi_{\pi + \frac14 \phi} \pi_{0} \pi_{ - \frac14 \phi} \pi_{\pi - \frac12 \phi}.
\end{align}
\ese
These four sequences generate the same propagator and hence the same fidelity.
The Frobenius distance and trace distance fidelities read
\bse
\begin{align}
& \mathcal{F} = 1 - \sqrt{2} \sin^2 \frac{\pi  \epsilon }{2}  \left| \sin \frac{\phi}{4} \right|,  \label{F-4-distance} \\
& \mathcal{F}_T = 1 - 2  \sin^4 \frac{\pi  \epsilon }{2} \sin^2 \frac{\phi }{4}.  \label{F-4-trace}
\end{align}
\ese
Obviously, the  Frobenius distance infidelity for four sequences is of order $O(\epsilon^2)$ and it is much larger than the trace distance infidelity, which is of order $O(\epsilon^4)$. The trace distance fidelity is much higher than the Frobenius distance fidelity, similar to rotation gates \cite{gevorgyan2021}.

For the Z, S and T gates of the form \eqref{phase-four-0} we have
\bse
\begin{align}
{Z4} &= \pi_{0} \pi_{- \frac14 \pi} \pi_{\frac12 \pi} \pi_{\frac14 \pi} , \\
{S4} &= \pi_{0} \pi_{- \frac18 \pi} \pi_{\frac34 \pi} \pi_{\frac58 \pi} , \\
{T4} &= \pi_{0} \pi_{- \frac1{16{}} \pi} \pi_{\frac78 \pi} \pi_{\frac{13}{16} \pi} .
\end{align}
\ese

The Frobenius distance fidelity $\mathcal{F}$ (top) and infidelity $1-\mathcal{F}$ (bottom) for the composite Z4, S4 and T4 gates is shown in Figs.~\ref{fig:Z}, \ref{fig:S}, and \ref{fig:T}, respectively.
With respect to the ubiquitous quantum computation benchmark fidelity value of $1-10^{-4}$, the Frobenius distance fidelity \eqref{F-4-distance} for the four-pulse composite Z4 gates of Eqs.~\eqref{phase4-two} remains above this value in the pulse area interval $(0.9936\pi, 1.0064\pi)$, i.e. for relative errors up to $|\epsilon| < 0.0064$. 
For comparison, the trace distance fidelity \eqref{F-4-trace} remains above this value in the pulse area interval $(0.936\pi, 1.064\pi)$, i.e. for relative errors up to $|\epsilon| < 0.064$, a factor of 10 larger.
Again we stress that the Frobenius distance fidelity is a much more stringent measure.

For the four-pulse composite S4 gate, the Frobenius infidelity remains below $10^{-4}$ for $|\epsilon| < 0.0087$, and the trace infidelity requires $|\epsilon| < 0.087$, a factor of 10 larger. 
For the four-pulse composite T4 gate, the Frobenius infidelity requires $|\epsilon| < 0.0121$, and the trace infidelity demands $|\epsilon| < 0.122$, a factor of 10 larger.

Obviously, both the Frobenius and trace distance fidelities depend on the phase flip angle $\phi$. 
The pulse area intervals for the four-pulse composite phase gates are broadest for T4 and narrowest for Z4, with S4 in the middle.
This is clearly visible in Figs.~\ref{fig:Z}, \ref{fig:S}, and \ref{fig:T}.
This monotonic pattern persists for longer sequences as well.

\subsection{Second-order error compensation}\label{Sec:phase-6}

For sequences of six $\pi$ pulses, it becomes possible to annul also the second-order derivatives in Eq. \eqref{eq-m-phase}. 
It is still possible to derive analytic solutions of the form
\be\label{phase6}
\pi_{\phi_0} \pi_{\phi_1} \pi_{\phi_2} \pi_{\phi_0 + \pi - \frac12 \phi} \pi_{\phi_1 + \pi - \frac12 \phi} \pi_{\phi_2 + \pi - \frac12 \phi},
\ee
The solutions can be written as
\bse \label{phase-six}
\begin{align} 
& \pi_{\chi} (2\pi)_{0} \pi_{\chi + \pi - \frac12 \phi} (2\pi)_{\pi - \frac12 \phi}, \label{phase-six-1} \\
& \pi_{\pi + \frac12 \phi - \chi} (2\pi)_{0} \pi_{-\chi} (2\pi)_{\pi - \frac12 \phi}, \\
& (2\pi)_{0} \pi_{\pi - \frac12 \phi + \chi} (2\pi)_{\pi - \frac12 \phi} \pi_{- \phi + \chi}, \\
& (2\pi)_{0} \pi_{- \chi} (2\pi)_{\pi - \frac12 \phi} \pi_{- \chi + \pi - \frac12 \phi}, \label{phase-six-0}
\end{align}
\ese
where 
\be
\chi = \frac{\phi}4 + \arcsin\left(\frac12 \sin\frac{\phi}4\right).
\ee
For the Z6 ($\phi=\pi$), S6 ($\phi=\pi/2$), and T6 ($\phi=\pi/4$) gates, we have $\chi=0.3650\pi$, $0.1863\pi$, and $0.0936\pi$, respectively.
Explicitly, for the Z, S and T gates of the form \eqref{phase-six-1} we have
\bse
\begin{align}
{Z6} &= \pi_{0.3650\pi} (2\pi)_0 \pi_{0.8650\pi} (2\pi)_{\frac12 \pi} , \\
{S6} &= \pi_{0.1863\pi} (2\pi)_0 \pi_{0.9363\pi} (2\pi)_{\frac34 \pi} , \\
{T6} &= \pi_{0.0936\pi} (2\pi)_0 \pi_{0.9686\pi} (2\pi)_{\frac78 \pi} .
\end{align}
\ese

The Frobenius distance and trace distance fidelities for these second-order sequences read
\bse
\begin{align}
& \mathcal{F} = 1 - \sqrt{2} \left| \sin^3 \frac{\pi  \epsilon }{2} \right| \left| \sin \frac{\phi}{4} \right|,  \label{F-6-distance} \\
& \mathcal{F}_T = 1 - 2  \sin^6 \frac{\pi  \epsilon }{2} \sin^2 \frac{\phi }{4}.  \label{F-6-trace}
\end{align}
\ese
The Frobenius infidelity is of order $O(\epsilon^3)$, and the trace infidelity is of order $O(\epsilon^6)$.
The Frobenius distance fidelity $\mathcal{F}$ (top) and infidelity $1-\mathcal{F}$ (bottom) for the composite Z6, S6 and T6 gates is shown in Figs.~\ref{fig:Z}, \ref{fig:S}, and \ref{fig:T}, respectively.

\subsection{Third-order error compensation}\label{Sec:phase-8}

Nullification of up to the third-order derivatives in Eq.~\eqref{eq-m-phase} requires eight $\pi$ pulses, 
\be\label{phase8}
\pi_{\phi_0} \pi_{\phi_1} \pi_{\phi_2} \pi_{\phi_3} \pi_{\phi_0 + \pi - \frac12 \phi} \pi_{\phi_1 + \pi - \frac12 \phi} \pi_{\phi_2 + \pi - \frac12 \phi} \pi_{\phi_3 + \pi - \frac12 \phi}.
\ee
The explicit solutions we have found are
\bse \label{phase-eight}
\begin{align} 
& \pi_{\chi} (2\pi)_{0} \pi_{\chi + \pi - \frac14 \phi} \pi_{\chi + \pi - \frac12 \phi} (2\pi)_{\pi - \frac12 \phi} \pi_{\chi - \frac34 \phi}, \label{phase-eight-1} \\
& \pi_{- \chi + \pi + \frac14 \phi} (2\pi)_{0} \pi_{- \chi} \pi_{- \chi - \frac14 \phi} (2\pi)_{\pi - \frac12 \phi} \pi_{- \chi + \pi - \frac12 \phi}, \\
& (2\pi)_{0} \pi_{\chi + \pi - \frac14 \phi} \pi_{\chi + \pi - \frac12 \phi} (2\pi)_{\pi - \frac12 \phi} \pi_{\chi - \frac34 \phi} \pi_{\chi - \phi}, \\
& (2\pi)_{0} \pi_{- \chi} \pi_{- \chi - \frac14 \phi} (2\pi)_{\pi - \frac12 \phi} \pi_{- \chi + \pi - \frac12 \phi} \pi_{- \chi + \pi - \frac34 \phi}, \label{phase-eight-0}\\
& \pi_{\chi + \frac14 \phi} \pi_{\chi} (2\pi)_{0} \pi_{\chi + \pi - \frac14 \phi} \pi_{\chi + \pi - \frac12 \phi} (2\pi)_{\pi - \frac12 \phi}, \\
& \pi_{- \chi + \pi + \frac12 \phi} \pi_{- \chi + \pi + \frac14 \phi} (2\pi)_{0} \pi_{- \chi} \pi_{- \chi - \frac14 \phi} (2\pi)_{\pi - \frac12 \phi},
\end{align}
\ese
where 
\be
\chi = \frac{\phi}8 + \arcsin\left(\frac12 \sin\frac{\phi}8 \right).
\ee
For the Z8 ($\phi=\pi$), S8 ($\phi=\pi/2$), and T8 ($\phi=\pi/4$) gates, we have $\chi=0.1863\pi$, $0.0936\pi$, and $0.0469\pi$, respectively.
Explicitly, for the Z, S and T gates of the form \eqref{phase-eight-1} we have
\bse
\begin{align}
{Z8} &= \pi_{0.1863\pi} (2\pi)_0 \pi_{0.9363\pi} \pi_{0.6863\pi} (2\pi)_{\frac12 \pi} \pi_{1.4363\pi} , \\
{S8} &= \pi_{0.0936\pi} (2\pi)_0 \pi_{0.9686\pi} \pi_{0.8436\pi} (2\pi)_{\frac34 \pi} \pi_{1.7186\pi} , \\
{T8} &= \pi_{0.0469\pi} (2\pi)_0 \pi_{0.9844\pi} \pi_{0.9219\pi} (2\pi)_{\frac78 \pi} \pi_{1.8594\pi} .
\end{align}
\ese

The Frobenius distance and trace distance fidelities for these third-order sequences read
\bse
\begin{align}
& \mathcal{F} = 1 - \sqrt{2} \sin^4 \frac{\pi  \epsilon }{2} \left| \sin \frac{\phi}{4} \right|,  \label{F-8-distance} \\
& \mathcal{F}_T = 1 - 2  \sin^8 \frac{\pi  \epsilon }{2} \sin^2 \frac{\phi }{4}.  \label{F-8-trace}
\end{align}
\ese
The Frobenius infidelity is of order $O(\epsilon^4)$, and the distance infidelity is of order $O(\epsilon^8)$.
The Frobenius distance fidelity $\mathcal{F}$ (top) and infidelity $1-\mathcal{F}$ (bottom) for the composite Z8, S8 and T8 gates is shown in Figs.~\ref{fig:Z}, \ref{fig:S}, and \ref{fig:T}, respectively.

\subsection{Higher-order error compensation}\label{Sec:phase-higher}

For CP sequences of more than eight $\pi$ pulses, the equations for the composite phases quickly get very bulky and hard to derive analytically. 
The general form for these sequences is given by Eq.~\eqref{phase-asymmetric-0}. 
%
They reiterate the pattern of the sequences of four, six and eight pulses above: the CP sequences of $2(n+1)$ $\pi$ pulses 
produce error compensation of the order $O(\epsilon^n)$ and fidelity profiles
\bse\label{F-higher}
\begin{align}
& \mathcal{F} = 
1 - \sqrt{2} \left| \sin^{n+1} \frac{\pi  \epsilon }{2} \right| \left| \sin \frac{\phi}{4} \right|,  \label{F-higher-distance} \\
& \mathcal{F}_T = 
1 - 2  \sin^{2n+2} \frac{\pi  \epsilon }{2} \sin^2 \frac{\phi }{4}.  \label{F-higher-trace}
\end{align}
\ese
The fidelities are sensitive to the choice of the gate phase $\phi$, as before.
These sequences are shown below.

We have derived numerically composite sequences consisting of only $\pi$ and $2\pi$ pulses. 
Other possible sequences are listed in Appendix \ref{Appendix:B}.
All composite sequences have the familiar structure of Eq.~\eqref{phase-asymmetric-0}, viz.
\be 
 R_{n+1} R_{n+1}(\pi-\phi/2), 
\ee
with
\bse
\begin{align}
R_5 &= \pi_{0} (2\pi)_{\phi_1} \pi_{\phi_3} \pi_{\phi_4} , \\
R_6 &= (2\pi)_{0} (2\pi)_{\phi_2} \pi_{\phi_4} \pi_{\phi_4 - \frac14\phi} , \\
R_7 &= \pi_{0} (2\pi)_{\phi_1} (2\pi)_{\phi_3} \pi_{\phi_5} \pi_{\phi_6} , \\
R_8 &= (2\pi)_{0} (2\pi)_{\phi_2} (2\pi)_{\phi_4} \pi_{\phi_6} \pi_{\phi_6 - \frac14 \phi} , \\
R_9 &= \pi_{0} (2\pi)_{\phi_1} (2\pi)_{\phi_3} (2\pi)_{\phi_5} \pi_{\phi_7} \pi_{\phi_8}.
\end{align}
\ese
They compensate errors of orders ranging from $O(\epsilon^4)$ for $R_5$ to $O(\epsilon^8)$ for $R_9$.
Other equivalent (in terms of total pulse area and fidelity) configurations can be obtained by interchanging pulses in the corresponding sequence similarly to Eqs.~\eqref{phase4-two}, \eqref{phase-six} and \eqref{phase-eight}.  

We have derived numerically the composite phases of this type of sequences of an even number of pulses.
They are presented in Tables~\ref{Table:Z},~\ref{Table:S} and~\ref{Table:T} for Z, S and T gates correspondingly.
The fidelities of these composite Z, S and T gates are plotted in Figures~\ref{fig:Z}, \ref{fig:S}, and \ref{fig:T} respectively.

\subsection{Discussion}

It can be seen from the tables and the figures that the two-pulse sequences Z2, S2 and T2 have very little room for errors, since high-fidelity Z, S and T gates allow pulse area errors of less than 0.01\%, about 0.01\%, about 0.02\%, respectively.
The four-pulse composite phase gates Z4, S4 and T4 offer some leeway, with the admissible error of 0.6\%, 0.9\% and 1.2\% for Z, S and T cases.
The significant pulse area error correction effect is achieved with the CP sequences of 6 to 10 pulses, for which the high-fidelity range of admissible errors increases from 3\% to 10.1\% for Z, from 3.6\% to 11.5\% for S, and from 4.5\% to 13.1\% for T.
Quite remarkably, errors of up to 23.4\%, 25.1\% and 27.1\% can be eliminated for Z, S and T, and ultrahigh fidelity maintained, with the 18-pulse composite phase gates Z18, S18 and T18.
Note that these error ranges are calculated by using the rather tough Frobenius distance fidelity \eqref{Frobenius}.
Had we used the much more relaxed trace distance fidelity \eqref{trace fidelity}, these ranges would be much broader. 

It is obvious from the discussion that by using longer composite sequences one can compensate increasingly large errors. 
However, very long sequences are barely practical because the gate is much slower.
Moreover, the quantum computer is not supposed to operate with a pulse area error of 10\% or more.
Clearly, there is some ``sweet spot'' of speed and error tolerance.
To this end, the CP sequences of 4 to 8 pulses, for which the phases are given by analytic formulas, seem to offer the best fidelity-to-speed ratio. 

Appendix \ref{Appendix:B} presents composite pulse sequences for general phase gates with different phase angles.

\section{Comments and conclusions\label{Sec:concl-ch3}}

In this paper we presented a number of CP sequences for four basic quantum gates --- the Z gate, the S gate, the T gate and general phase gates.
The CP sequences contain up to 18 pulses and can compensate up to eight orders of experimental errors in the pulse amplitude and duration.
The short CP sequences (up to 8 pulses) are calculated analytically and the longer ones numerically.
Although longer composite phase gates are derived numerically, their fidelity profiles have an analytic dependence on the pulse area error, Eqs.~\eqref{F-higher-distance} and \eqref{F-higher-trace}, and show trigonometric dependence on the phase-shift angle. 

A similar class of CP sequences for phase gates is derived in \cite{torosov2014}, where they are build from the $\theta$ rotation gates, containing $2n+1$ ($n=1,2,\ldots$) pulses, in a similar scenario as prescribed by Eq.~\eqref{phase-asymmetric-0}.
Hence the composite sequences contain $2(2n+1)$ pulses, i.e. 6, 10, 14, ... pulses.
Here we fill the missing numbers of 4, 8, 12, ... pulses, of which particularly important appear to be the 4-pulse sequences, as they are the fastest ones. 
The sequences with the same number of pulses here and in Ref.~\cite{torosov2014}, although different in construction, have performance equal to the earlier composite gates. 


The results presented in this article demonstrate the remarkable flexibility of CPs accompanied by extreme accuracy and robustness to errors --- three features that cannot be achieved together by any other coherent control technique.
We expect these CP sequences, in particular the Z, the S and the T gates, to be very useful quantum control tools in quantum computing applications, because they provide a variety of options to find the optimal balance between ultrahigh fidelity, error range and speed, which may be different in different physical applications.

We note that in addition to quantum computing, the results presented in this paper can be applied in polarization optics to obtain broadband polarization rotators using stacked single polarization half-wave plates with the optical axes rotated by precisely chosen rotation angles (composite phases). 
This is possible due to quantum-classical analogy of composite rotations on the Bloch and the Poincar\'e spheres \cite{rangelov}. 
Hereby, we demonstrate the possibility to design the broadband polarization rotators with $\pi/2$, $\pi/4$, $\pi/8$ and arbitrary phase shift angles, by up to 18 CP sequences.



\acknowledgments
HLG acknowledges support from the EU Horizon-2020 ITN project LIMQUET (Contract No. 765075), and also from the RA Science Committee in the frames of the research project 20TTAT-QTc004.
NVV acknowledges support from the Bulgarian national plan for recovery and resilience, contract BG-RRP-2.004-0008-C01 (SUMMIT), project number 3.1.4.

\appendix

\section{Composite phases for Z, S and T phase gates}

Here we present the complete sets of phases of the composite pulse sequences generating phase gates with various orders of error compensation.
\begin{table*}[tbph]
\begin{tabular}{|c|c|c|l|c|}
\hline
{\bfseries Name } & {\bfseries Pulses } &{\bfseries $O(\epsilon^n)$ } & {\bfseries Phases $\phi_0,\phi_1,\phi_2, \ldots, \phi_n, \phi_{n+1}, \ldots, \phi_{2n+1}$} (in units $\pi$) & {\bfseries High-fidelity } \\
& & & (according to \eqref{phase-asymmetric}) & {\bfseries error correction range } \\
\hline
two & 2 & $O(\epsilon^0)$ & $0, \frac12$ & $ [0.99994\pi, 1.00006\pi] $ \\

Z4 & 4 & $O(\epsilon)$ & $0, \frac74, \frac12, \frac14$ & $ [0.994\pi, 1.006\pi] $ \\
Z6 & 6 & $O(\epsilon^2)$ & $0, 0, 1.6350, \frac12, \frac12, 0.1350$ & $ [0.970\pi, 1.030\pi] $ \\

Z8 & 8 & $O(\epsilon^3)$ & $0, 0, 1.8137, 1.5637, \frac12, \frac12, 0.3137, 0.0637 $ & $ [0.936\pi, 1.064\pi] $ \\
Z10 & 10 & $O(\epsilon^4)$ & $0, 1.0992, 1.0992, 1.8315,
0.0203, \frac12, 1.5992, 1.5992, 0.3315, 0.5203$  & $ [0.899\pi, 1.101\pi] $ \\

Z12 & 12 & $O(\epsilon^5)$ & $0, 0, 0.4492, 0.4492, 1.4099, 1.1599, \frac12, \frac12, 0.9492, 0.9492, 1.9099, 1.6599$ & $ [0.862\pi, 1.138\pi] $ \\
Z14 & 14 & $O(\epsilon^6)$ & $0, 0.7815, 0.7815, 1.9963, 1.9963, 0.8915, 0.3245, \frac12, 1.2815, 1.2815, 0.4963,$ & \\
 & & & \quad $0.4963, 1.3915, 0.8245$ & $[0.823\pi, 1.177\pi] $ \\
 
Z16 & 16 & $O(\epsilon^7)$ & $0, 0, 1.8969, 1.8969, 1.0586, 1.0586, 0.0214, 1.7714, \frac12, \frac12, 0.3969, 0.3969, 1.5586,$ & \\

 & & & \quad $1.5586, 0.5214, 0.2714$ & $[0.795\pi, 1.205\pi]$ \\
Z18 & 18 & $O(\epsilon^8)$ & $0, 0.1421, 0.1421, 1.0834, 1.0834, 0.5572, 0.5572, 1.4991, 1.0352, \frac12, 0.6421, 0.6421,$ & \\
 & & & \quad $1.5834, 1.5834, 1.0572, 1.0572, 1.9991, 1.5352$ & $[0.766\pi, 1.234\pi]$ \\
\hline
\end{tabular}
\caption{
Phases of asymmetric composite sequences of $N=2(n+1)$ nominal $\pi$ pulses, which produce the Z gate with a pulse area error compensation up to order $O(\epsilon^n)$.
The last column gives the high-fidelity range $[\pi (1-\epsilon_0), \pi (1+\epsilon_0)]$ of pulse area error compensation wherein the Frobenius distance fidelity is above the value $0.9999$, i.e. the fidelity error is below $10^{-4}$.}
\label{Table:Z}
\end{table*}

\begin{table*}
\begin{tabular}{|c|c|c|l|c|}
\hline
{\bfseries Name } & {\bfseries Pulses } &{\bfseries $O(\epsilon^n)$ } & {\bfseries Phases $\phi_0,\phi_1,\phi_2, \ldots, \phi_n, \phi_{n+1}, \ldots, \phi_{2n+1}$} (in units $\pi$) & {\bfseries High-fidelity } \\
  &  &  & (according to \eqref{phase-asymmetric}) & {\bfseries error correction range } \\
\hline
two & 2 & $O(\epsilon^0)$ & $0, \frac34$ & $ [0.99988\pi, 1.00012\pi] $ \\

S4 & 4 & $O(\epsilon)$ & $0, \frac{15}{8}, \frac34, \frac58$ & $ [0.991\pi, 1.009\pi] $ \\
S6 & 6 & $O(\epsilon^2)$ & $0, 0, 1.8137, \frac34, \frac34, 0.5637$ & $ [0.964\pi, 1.036\pi] $ \\

S8 & 8 & $O(\epsilon^3)$ & $0, 0, 1.9064, 1.7814, \frac34, \frac34, 0.6564, 0.5314 $ & $ [0.926\pi, 1.074\pi] $ \\
S10 & 10 & $O(\epsilon^4)$ & $0, 0.8226, 0.8226, 1.9152, 0.4416, \frac34, 1.5726, 1.5726, 0.6652, 1.1916$  & $ [0.885\pi, 1.115\pi] $ \\

S12 & 12 & $O(\epsilon^5)$ & $0, 0, 1.3587, 1.3587, 0.3367, 0.2117, \frac34, \frac34, 0.1087, 0.1087, 1.0867, 0.9617$ & $ [0.847\pi, 1.153\pi] $ \\
S14 & 14 & $O(\epsilon^6)$ & $0, 0.8197, 0.8197, 1.6756, 1.6756, 0.7586, 1.1000, \frac34, 1.5697, 1.5697, 0.4255,$ & \\
 & & & \quad $0.4255, 1.5086, 1.8500$ & $[0.811\pi, 1.189\pi]$ \\
 
S16 & 16 & $O(\epsilon^7)$ & $0, 0, 1.9466, 1.9466, 1.1420, 1.1420, 0.1251, 0.0001, \frac34, \frac34, 0.6966, 0.6966, 1.8920,$ & \\

 & & & \quad $1.8920, 0.8751, 0.7501$ & $[0.778\pi, 1.222\pi]$ \\
S18 & 18 & $O(\epsilon^8)$ & $0, 0.3453, 0.3453, 1.4636, 1.4636, 0.2616, 0.2616, 1.3543, 0.0643, \frac34, 1.0953, 1.0953,$ & \\
 & & & \quad $0.2136, 0.2136, 1.0116, 1.0116, 0.1043, 0.8143$ & $[0.749\pi, 1.251\pi]$ \\
\hline
\end{tabular}
\caption{
Phases of asymmetric composite sequences of $N=2n+2$ nominal $\pi$ pulses, which produce the S gate with a pulse area error compensation up to order $O(\epsilon^n)$.
The last column gives the high-fidelity range $[\pi (1-\epsilon_0), \pi (1+\epsilon_0)]$ of pulse area error compensation wherein the Frobenius distance fidelity is above the value $0.9999$, i.e. the fidelity error is below $10^{-4}$.}
\label{Table:S}
\end{table*}

\begin{table*}
\begin{tabular}{|c|c|c|l|c|}
\hline
{\bfseries Name } & {\bfseries Pulses } & {\bfseries $O(\epsilon^n)$ } & {\bfseries Phases $\phi_0,\phi_1,\phi_2, \ldots, \phi_n, \phi_{n+1}, \ldots, \phi_{2n+1}$} (in units $\pi$) & {\bfseries High-fidelity} \\
 &  &  & (according to \eqref{phase-asymmetric}) & {\bfseries error correction range } \\
\hline
two & 2 & $O(\epsilon^0)$ & $0, \frac78$ & $ [0.99977\pi, 1.00023\pi] $ \\

T4 & 4 & $O(\epsilon)$ & $0, \frac{31}{16}, \frac78, \frac{13}{16}$ & $ [0.988\pi, 1.012\pi] $ \\
T6 & 6 & $O(\epsilon^2)$ & $0, 0, 1.9064, \frac78, \frac78, 0.7814$ & $ [0.955\pi, 1.045\pi] $ \\

T8 & 8 & $O(\epsilon^3)$ & $0, 0, 1.9531, 1.8906, \frac78, \frac78, 0.8281, 0.7656 $ & $ [0.912\pi, 1.088\pi] $ \\
T10 & 10 & $O(\epsilon^4)$ & $0, 1.1086, 1.1086, 0.0218, 0.2593, \frac78, 1.9836, 1.9836, 0.8968, 1.1343$  & $[0.869\pi, 1.131\pi]$ \\

T12 & 12 & $O(\epsilon^5)$ & $0, 0, 0.5488, 0.5488, 1.5386, 1.4761, \frac78, \frac78, 1.4238, 1.4238, 0.4136, 0.3511$ & $[0.828\pi, 1.172\pi]$ \\
T14 & 14 & $O(\epsilon^6)$ & $0, 0.9406, 0.9406, 0.2214, 0.2214, 1.1532, 1.3379, \frac78, 1.8156, 1.8156, 1.0964,$ & \\
 & & & \quad $1.0964, 0.0282, 0.2129$ & $[0.791\pi, 1.209\pi]$ \\
 
T16 & 16 & $O(\epsilon^7)$ & $0, 0, 1.9724, 1.9724, 0.7247, 0.7247, 1.7171, 1.6546, \frac78, \frac78, 0.8474, 0.8474, 1.5997,$ & \\

 & & & \quad $1.5997, 0.5921, 0.5296$ & $[0.758\pi, 1.242\pi]$ \\
T18 & 18 & $O(\epsilon^8)$ & $0, 0.9424, 0.9424, 0.5711, 0.5711, 1.3429, 1.3429, 0.3645, 0.6381, \frac78, 1.8174, 1.8174,$ & \\
 & & & \quad $1.4461, 1.4461, 0.2179, 0.2179, 1.2395, 1.5131$ & $[0.728\pi, 1.272\pi]$ \\
\hline
\end{tabular}
\caption{
Phases of asymmetric composite sequences of $N=2n+2$ nominal $\pi$ pulses, which produce the T gate with a pulse area error compensation up to order $O(\epsilon^n)$.
The last column gives the high-fidelity range $[\pi (1-\epsilon_0), \pi (1+\epsilon_0)]$ of pulse area error compensation wherein the Frobenius distance fidelity is above the value $0.9999$, i.e. the fidelity error is below $10^{-4}$.}
\label{Table:T}
\end{table*}

\section{Arbitrary phase gates\label{Appendix:B}}

The fourth-order compensating sequences of ten $\pi$ pulses can be written in the compact form
\be\label{phase10-0}
(3\pi)_{0} \pi_{\phi_3} \pi_{\phi_4} (3\pi)_{\pi-\frac12\phi} \pi_{\phi_3+\pi-\frac12\phi} \pi_{\phi_4+\pi-\frac12\phi}.
\ee
For brevity, we do not show other configurations consisting of $3\pi$ pulse and $\pi$ pulses, because all these designs have equal total pulse area, i.e. operation run-time, and equal fidelity. 
Such solutions can be obtained by interchanging pulses in the sequence similar to \eqref{phase4-two}, \eqref{phase-six} and \eqref{phase-eight}.     

The fifth-order compensating sequence consists of twelve $\pi$ pulses and it can be written in the compact form
\be\label{phase12-0}
 (3\pi)_{0} \pi_{\phi_3} \pi_{\phi_4} \pi_{\phi_5} 
 (3\pi)_{\pi-\frac12\phi} \pi_{\phi_3+\pi-\frac12\phi}  \pi_{\phi_4+\pi-\frac12\phi} \pi_{\phi_5+\pi-\frac12\phi},
\ee
with $\phi_5 = \phi_4 - \phi_3 - \frac14\phi$.
The sixth-order compensating sequence contains fourteen $\pi$ pulses, 
\be\label{phase14-0}
(4\pi)_{0} \pi_{\phi_4} \pi_{\phi_5} \pi_{\phi_6} (4\pi)_{\pi-\frac12\phi} \pi_{\phi_3+\pi-\frac12\phi}  \pi_{\phi_4+\pi-\frac12\phi} \pi_{\phi_5 + \pi - \frac34\phi}.
\ee

The composite phases for this type of composite phase gates for arbitrary phase flip angles are presented in Table~\ref{Table:Phase}. 
The structure of these sequences corresponds to Eq.~\eqref{phase-asymmetric} with $\nu = 0$ and zero first phases, i.e. with accordance to Eqs.~\eqref{phase-four-0}, \eqref{phase-six-0}, \eqref{phase-eight-0}, \eqref{phase10-0}, \eqref{phase12-0} and \eqref{phase14-0}. 

\begin{table*}
\begin{tabular}{|c|c|c|c|c|c|c|}
\hline
{\bfseries } & {\bfseries 4 pulses, $O(\epsilon)$ } &{\bfseries 6 pulses, $O(\epsilon^2)$ } & {\bfseries 8 pulses, $O(\epsilon^3)$ } & {\bfseries 10 pulses, $O(\epsilon^4)$ } & {\bfseries 12 pulses, $O(\epsilon^5)$ } & {\bfseries 14 pulses, $O(\epsilon^6)$ } \\
 \hline
{\bfseries $\phi$ } & {\bfseries $\phi_1$ } &
{\bfseries $\phi_2$ } & {\bfseries $\phi_2, \phi_3$ } & {\bfseries $\phi_3, \phi_4$ } & {\bfseries $\phi_3, \phi_4, \phi_5$ } & {\bfseries $\phi_4, \phi_5, \phi_6$ } \\
& (see \eqref{phase-four-0}) & (see \eqref{phase-six-0}) & (see \eqref{phase-eight-0}) & (see \eqref{phase10-0}) & (see \eqref{phase12-0}) & (see \eqref{phase14-0}) \\
\hline
$\frac1{16}$ & \scriptsize $\frac{127}{64} = 1.984375$ & \scriptsize 1.9766 & \scriptsize 1.9883, 1.9727 & \scriptsize 0.9980, 0.9883 & \scriptsize 1.0316, 1.7227, 0.6755 & \scriptsize 0.9995, 0.9960, 0.9857 \\

$\frac1{12}$ & \scriptsize $\frac{95}{48} = 1.9791(6)$ & \scriptsize 1.9688 & \scriptsize 1.9844, 1.9635 & \scriptsize 0.9974, 0.9844 & \scriptsize 1.0342, 1.6996, 0.6446 & \scriptsize 0.9993, 0.9947, 0.9809 \\
$\frac18$ & \scriptsize $\frac{63}{32} = 1.96875$ & \scriptsize 1.9531 & \scriptsize 1.9766, 1.9453 & \scriptsize 0.9961, 0.9765 & \scriptsize 1.0379, 1.6646, 0.5955 & \scriptsize 0.9990, 0.9922, 0.9716 \\

$\frac16$ & \scriptsize $\frac{47}{24} = 1.958(3)$ & \scriptsize 1.9375 & \scriptsize 1.9688, 1.9271 & \scriptsize 0.9948, 0.9687 & \scriptsize 1.0405, 1.6378, 0.5556 & \scriptsize 0.9987, 0.9895, 0.9620 \\
$\frac14$ & \scriptsize $\frac{31}{16} = 1.9375$ & \scriptsize 1.9064 & \scriptsize 1.9531, 1.8906 & \scriptsize 0.9922, 0.9530 & \scriptsize 1.0439, 1.5966, 0.4901 & \scriptsize 0.9980, 0.9843, 0.9431 \\

$\frac13$ & \scriptsize $\frac{23}{12} = 1.91(6)$ & \scriptsize 1.8754 & \scriptsize 1.9375, 1.8542 & \scriptsize 0.9895, 0.9371 & \scriptsize 1.0459, 1.5642, 0.4349 & \scriptsize 0.9974, 0.9790, 0.9240 \\
$\frac12$ & \scriptsize $\frac{15}{8} = 1.875$ & \scriptsize 1.8137 & \scriptsize 1.9064, 1.7814 & \scriptsize 0.9842, 0.9050 & \scriptsize 1.0477, 1.5126, 0.3399 & \scriptsize 0.9961, 0.9684, 0.8855 \\

$\frac23$ & \scriptsize $\frac{11}{6} = 1.8(3)$ & \scriptsize 1.7529 & \scriptsize 1.8754, 1.7087 & \scriptsize 0.9787, 0.8721 & \scriptsize 1.0479, 1.4703, 0.2558 & \scriptsize 0.9947, 0.9575, 0.8460 \\
$\frac34$ & \scriptsize $\frac{29}{16} = 1.8125$ & \scriptsize 1.7229 & \scriptsize 1.8599, 1.6724 & \scriptsize 0.9759, 0.8552 & \scriptsize 1.0475, 1.4512, 0.2162 & \scriptsize 0.9941, 0.9520, 0.8259 \\

$\frac56$ & \scriptsize $\frac{43}{24} = 1.791(6)$ & \scriptsize 1.6932 & \scriptsize 1.8444, 1.6361 & \scriptsize 0.9731, 0.8381 & \scriptsize 1.0470, 1.4332, 0.1779 & \scriptsize 0.9934, 0.9464, 0.8056 \\
$\frac78$ & \scriptsize $\frac{57}{32} = 1.78125$ & \scriptsize 1.6785 & \scriptsize 1.8368, 1.6180 & \scriptsize 0.9717, 0.8294 & \scriptsize 1.0467, 1.4245, 0.1591 & \scriptsize 0.9930, 0.9436, 0.7953 \\

$\frac{11}{12}$ & \scriptsize $\frac{85}{48} = 1.7708(3)$ & \scriptsize 1.6639 & \scriptsize 1.8291, 1.5999 & \scriptsize 0.9702, 0.8206 & \scriptsize 1.0463, 1.4161, 0.1405 & \scriptsize 0.9927, 0.9407, 0.7849 \\
$\frac{15}{16}$ & \scriptsize $\frac{113}{64} = 1.765625$ & \scriptsize 1.6566 & \scriptsize 1.8252, 1.5908 & \scriptsize 0.9695, 0.8161 & \scriptsize 1.0462, 1.4119, 0.1314 & \scriptsize 0.9925, 0.9393, 0.7797 \\

$1$ & \scriptsize $\frac{7}{4} = 1.75$ & \scriptsize 1.6350 & \scriptsize 1.8137, 1.5637 & \scriptsize 0.9673, 0.8027 & \scriptsize 1.0456, 1.4000, 0.1041 & \scriptsize 0.9920, 0.9350, 0.7638 \\
\hline
\end{tabular}
\caption{
Phases of composite pulse sequences which produce phase gates of angle $\phi$ (according to \eqref{phase-asymmetric}).
The all phases are given in units $\pi$.
The cases of $\phi = \pi$, $\phi = \frac12 \pi$ and $\phi = \frac14 \pi$ repeat the asymmetric Z, S and T gates respectively, already presented partly in Sec.~\ref{Sec:phase_gates}, and partly in Appendix \ref{Appendix:B}; they are given here for the sake of comparison and completeness.}
\label{Table:Phase}
\end{table*}



\end{document}